\documentclass[letter]{aa}
\usepackage{graphicx}
\usepackage{txfonts}
\usepackage{booktabs}
\usepackage{hyperref} 

\begin{document}

   \title{Magnetic fields in planetary nebulae detected through non-thermal radio continuum emission}

   \author{Marcin Hajduk\inst{1}
          \and
          Timothy Shimwell\inst{2,3}
          \and
          Glenn White\inst{4,5}
          \and
          Marijke Haverkorn\inst{6}
          \and
          Jes\'{u}s A. Toal\'{a}\inst{7}
          \and
          Ralf-J\"{u}rgen Dettmar\inst{8}
          }

   \institute{Department of Geodesy, Faculty of Geoengineering, University of Warmia and Mazury, ul. Oczapowskiego 2,
    10-719 Olsztyn, Poland\\
              \email{marcin.hajduk@uwm.edu.pl}
         \and
             Leiden Observatory, Leiden University, PO Box 9513 2300 RA Leiden, The Netherlands
            \and
            ASTRON, The Netherlands Institute for Radio Astronomy, Postbus 2 NL-7990 AA Dwingeloo, The Netherlands
        \and
        School of Physical Sciences, The Open University, Walton Hall, Milton Keynes, MK7 6AA, UK
        \and
        RAL Space, STFC Rutherford Appleton Laboratory, Chilton, Didcot, Oxfordshire, OX11 0QX, UK
        \and
        Department of Astrophysics/IMAPP, Radboud University, PO Box 9010, 6500 GL Nijmegen, The Netherlands
        \and
        Instituto de Radioastronom\'{i}a y Astrof\'{i}sica, Universidad Nacional Aut\'{o}noma de M\'{e}xico, Morelia 58089, Mich., Mexico
        \and Ruhr University Bochum, Faculty of Physics and Astronomy, Astronomical Institute (AIRUB), 44780 Bochum, Germany
             }

   \date{Received September 15, 1996; accepted March 16, 1997}

  \abstract
   {Planetary nebulae are shells ejected by low- and intermediate-mass stars. The slow wind ejected by the asymptotic giant branch star is compressed by a fast stellar wind to produce an expanding gaseous shell surrounding a hot bubble. The shell is a source of thermal radio emission which shows a spectral index between $-0.1$ and $2$. Only two planetary nebulae are known to show non-thermal radio emission indicating magnetic fields and non-thermal electrons.
}
   {The aim of this paper is verification of presence of magnetic fields of planetary nebulae. Magnetic fields can have a significant influence on shaping planetary nebulae.}
   {We observed a sample of northern planetary nebulae in radio continuum at 144\,MHz with the Low Frequency Array. We combined our observations with archival observations at higher frequencies.}
   {The spectral indices in 30 planetary nebulae were below $-0.1$, indicating non-thermal radio emission. The majority of this sample consists of bipolar planetary nebulae, which are known to originate from binary central stars. Most of the nebulae have sizes larger than 20 arcsec. Magnetic fields and nonthermal emission may be common in smaller planetary nebulae, but can be suppressed by thermal emission. Our results suggest that different mechanisms can be responsible for the origin of magnetic fields and non-thermal emission in planetary nebulae.}
   {}


   \maketitle

\section{Introduction}

The formation and evolution of planetary nebulae (PNe) is tightly correlated to the dramatic changes experienced by their progenitor low- and intermediate-mass stars. When the star evolves into the asymptotic giant branch (AGB) phase, it exhibits a dense but slow wind \citep[$\dot{M}\lesssim10^{-5}$~M$_\odot$~yr$^{-1}$, $v\approx 15$~km~s$^{-1}$;][]{2005ARA&A..43..435H}. After a few $10^{5}$~years, the star can deposit up to 90\% of its initial mass into the interstellar medium, forming a dense, dust-rich shell. As a result of the powerful mass loss, the hot core of the star is exposed, becoming a post-AGB star. The new hot star then produces a strong UV flux and a fast wind \citep[$\approx$500--4000~km~s$^{-1}$;][]{GuerreroDeMarco2013} that together compress and ionize the previously ejected material, ultimately forming a PN \citep{2000oepn.book.....K}. 

PNe emit in a broad range of electromagnetic spectrum, from X-rays to radio wavelengths. Each wavelength range provides complementary information about the structure and evolution of PNe. In radio frequencies, the free-free thermal emission dominates at GHz frequencies and it provides information on ionized content of PNe, complementary to emission lines in optical and UV. Optically thin radio flux depends on the emission measure and has a spectral index of $-0.1$. Optically-thick emission depends on the electron temperature and produces a spectral index of $+2.0$. Partially optically thick plasma produces a spectral index between $-0.1$ and $2$. The spectral index below $-0.1$ indicates non-thermal emission, or a mixture of thermal and non-thermal emission \citep{2018MNRAS.479.5657H}.

Non-thermal radio emission is a signature of magnetic fields and energetic electrons. Non-thermal emission arising from wind-shock interactions was observed in a few proto-PNe \citep{2009MNRAS.397.1386B}. It was searched unsuccessfully in PNe since the 1960s until \citet{2015ApJ...806..105S} found nonthermal emission in IRAS\,15103-5754. A few years later, \citet{2024A&A...688L..21H} reported non-thermal emission in a born-again PN Sakurai's Object at cm wavelengths. \citet{2017MNRAS.468.3450C} list a few more candidates that may show non-thermal emission.

Magnetic fields can play an important role in shaping planetary nebulae \citep{2001Natur.409..485B}. Magnetic fields have been suggested to be a necessary ingredient in the common envelope scenario, which is responsible for the formation of bipolar PNe \citep{2012ApJ...761..172G}. The lack of radio continuum non-thermal emission from PNe could indicate that magnetic fields do not exist in the majority of PNe, synchrotron emission is too faint compared to thermal emission, or it is embedded in optically thick thermal emission \citep{2018MNRAS.479.5657H}. Optical depth of thermal emission increases with $\nu^{-2}$, which may make the detection of synchrotron emission problematic at low frequencies. Non-thermal emission was not observed at low frequencies yet  \citep{2021ApJ...919..121H,2025PASA...42..111A}. 

Good spatial resolution and sensitivity could help us to observe non thermal emission at low frequencies \citep{1998ApJ...499L..83D}. In this paper, we report detection of synchrotron emission in a sample of PNe using Low-Frequency Array \citep[LOFAR;][]{2013A&A...556A...2V}.

\section{Identification of planetary nebulae at 144\,MHz}

LOFAR is the first interferometer operating in the MHz regime on very long baselines, including 54 stations distributed over Europe. Each station can operate independently or within the International LOFAR Telescope (ILT) network with High Band Antennas (HBA) observing at 120--240 MHz or Low Band Antennas (LBA) observing at 10--90 MHz.

One of the key projects is the LOFAR Two-metre Sky Survey (LoTSS)  \citep{2022A&A...659A...1S}. Radio maps at 128--160\,MHz (average frequency of 144\,MHz) have been restored with a 6 arcsec beam. The spatial resolution and average sensitivity of 100~$\mu$Jy are much deeper than previous surveys in this spectral range. We used LOFAR Data Release 3 (DR 3) observations, which cover about 80\% of the northern sky.

We selected 600 true, possible, and likely PNe within the LOFAR DR3 fields using The University of Hong Kong/Australian Astronomical Observatory/Strasbourg Observatory H-alpha Planetary Nebula (HASH PN) database of Galactic planetary nebulae \citep{2016JPhCS.728c2008P}. For relatively compact and coincident sources cross-matching between LOFAR and optical images would be relatively straightforward \citep{2019A&A...622A...2W}. However, we could not employ classical methods for positional matching in our case because most of the objects are extended and their positional precision is not well defined, and the surface brightness distribution can be different in $\mathrm{H}\alpha$ and at 144\,MHz. 

HASH positions are measured from aperture and centroid fitting of $\mathrm{H}\alpha$ emission \citep{2016JPhCS.728c2008P}. The positional accuracy is not given. The sources are detected in LOFAR-DR3 images with the Python Blob Detection and Source Finder (PyBDSF) \citep{2015ascl.soft02007M} and fitted with a Gaussian. The positional accuracy ranges from about 0.2\,arcsec to 0.5\,arcsec for high and low S/N sources, respectively. The positional accuracy is worse for extended sources.

For optically thin thermal emission at 144\,MHz, the brightness distribution reflects primarily emission measure in different directions, similar to $\mathrm{H}\alpha$. Surface brightness distribution of a nebula optically thick at 144\,MHz depends on $\mathrm{T}_\mathrm{e}$ and would be flat for a PN with homogeneous electron temperature. The regions where magnetic fields and shocks producing non-thermal electrons exist may be the source of non-thermal radio emission, which brightness distribution may be significantly different from $\mathrm{H}\alpha$ emission.

\begin{figure}[h]
\centering
\includegraphics[width=\columnwidth]{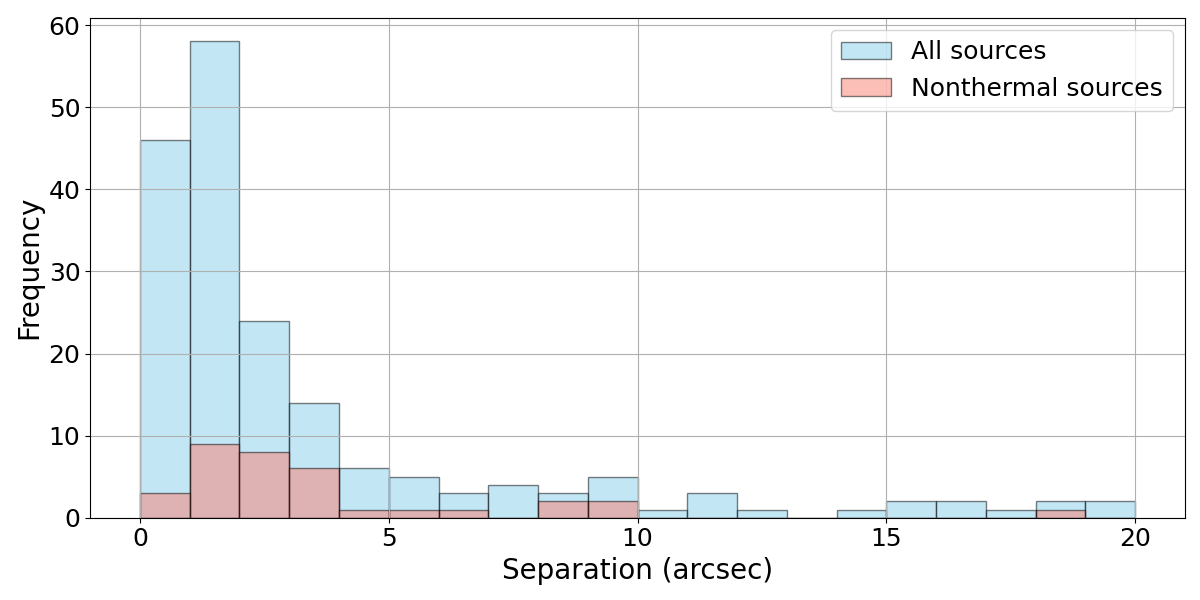}
\caption{Separations of the PNe and LoTSS DR3 sources.}\label{fig1}
\end{figure}

We cross-matched positions of PNe with the positions of radio sources in the LOFAR-DR3 source catalog. The initial matching radius was 60 arcsec. It resulted in 257 matches. The histogram of source separations reveals a maximum corresponding to a separation of about 2 arcsec with a tail resulting from variable position precision \citep{2019RNAAS...3...37C} and some background objects (Figure\,\ref{fig1}). The separation of the source counterparts can be larger in the case of different surface brightness distribution in $\mathrm{H}\alpha$ and 144\,MHz. Adopting the mean positional accuracy of PNe of 2 arcsec, corresponding roughly to the width of the peak in the histogram, we would expect less than 1 false positive (i.e., background object appearing closer than 2 arcsec to a PN) in the whole sample.

We combined LoTSS data with observations at higher frequencies \citep{1989A&AS...79..329Z,1991ApJS...75.1011G,1996ApJS..103..427G,1997A&AS..124..259R,1998AJ....115.1693C,2015MNRAS.453.1396P,2021ApJS..255...30G,2021PASA...38...58H,2025PASA...42...38D} to derive spectral indices and determine the radio emission mechanism. We used both interferometric surveys and single-dish surveys to improve frequency coverage. \citet{2018MNRAS.479.5657H} showed good agreement between flux densities of the same PNe measured using single-dish antennas and interferometers. However, single-dish observations are more vulnerable to blending with background objects due to worse angular resolution. We inspected radio maps and removed single dish measurements blended with background sources.

The fluxes of the objects can be affected by the limited sensitivity of the interferometer to extended emission. This can occur in interferometers with a limited number of short baselines. LOFAR has a dense core that enables us to observe large structures extending to a few degrees in the sky. However, we removed data points where the source size exceeded the largest angular scale for the VLASS survey (53\,arcsec).

We calculated spectral indices with a linear fit to all datapoints (Figure\,A2, A3 and A4). The non-thermal sources having spectral indices $<-0.1$ are plotted with red bars in the histogram in Figure\,{\ref{fig1}} and the remaining sources are plotted with blue bars. The distribution of non-thermal sources shows a maximum, which would not be expected in the case of a random distribution. Thus, at least some of the non-thermal sources must be real counterparts of the PNe. We inspected each case individually to remove the remaining background galaxies. We assumed the cases where a compact radio source was located within an extended PN but without any link to the PN morphology to be a background object. We identified 30 PNe showing non-thermal radio emission. Their properties are listed in Table\,A1. We did not include objects which did not have counterparts in other radio surveys, so that we could not verify their emission mechanism.

\begin{figure*}[h]
\centering
\includegraphics[width=0.33\textwidth]{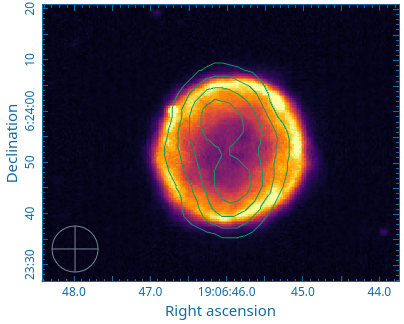}\includegraphics[width=0.33\textwidth]{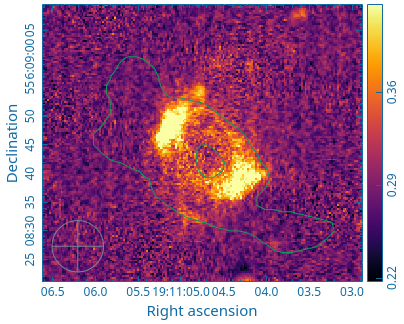}\includegraphics[width=0.33\textwidth]{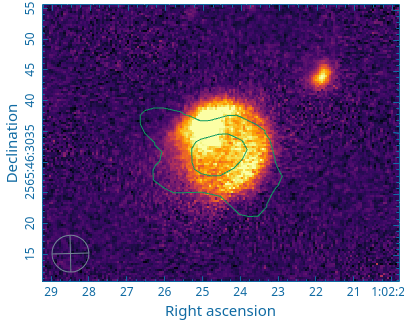}
\includegraphics[width=0.33\textwidth]{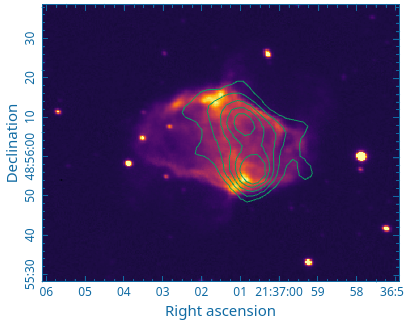}
\includegraphics[width=0.33\textwidth]{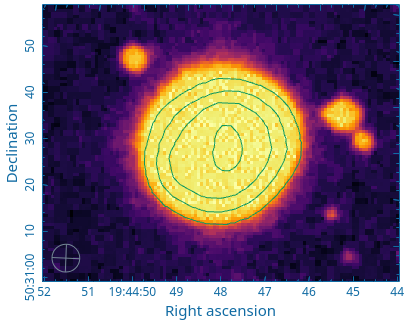}\includegraphics[width=0.33\textwidth]{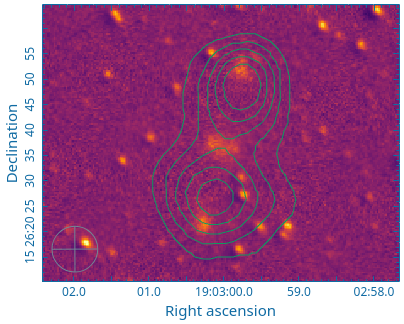}
\caption{Examples of 144\,MHz radio (contours) and optical (background) emission: Abell\,53, IRAS\,19086+0603, KLSS\,2-7, M\,1-79, NGC\,6826, PaEk\,1. The beam size is indicated by a circle in the bottom left corner.}\label{images}
\end{figure*}

\section{Non-thermal emission in PNe}

Non-thermal emission and magnetic fields in PNe may be generated with a few different mechanisms. Bipolar non-thermal emission can be explained by winds from magnetized central stars \citep{1994ApJ...421..225C}. In such case synchrotron emission is concentrated towards the polar axis of the central star. 

The intersection of hot bubbles with the slower shell can raise shocks with temperatures reaching $1-3\,\mathrm{MK}$ \citep{2000ApJ...545L..57K} in which non-thermal electrons are produced. Magnetic fields can be amplified in the shocked hot bubble \citep{1994ApJ...421..225C}.

Radio non-thermal emission may originate from the interaction of a magnetized and fast wind with dense knots in the PN \citep{1998ApJ...499L..83D,2007MNRAS.382.1607C}. This can produce a patchy emission, which may be observed if radio images have sufficient angular resolution.
 
An important clue in determining the origin of non-thermal radio emission is the surface brightness distribution of radio continuum emission, and the comparison of the optical and radio images. We plotted non-thermal 144\,MHz emission and $\mathrm{H}\alpha$ emission for a few representative cases in Figure\,\ref{images}. $\mathrm{H}\alpha$ and 144\,MHz emission for all PNe showing non-thermal emission is plotted in Figures\,A4, A5, A6 and A7. 

In several cases, non-thermal radio emission in PNe fills low surface-brightness regions in $\mathrm{H}\alpha$. This could be the case for NGC\,6781 and NGC\,6894 (Figure\,A6). The 144\,MHz sizes of these PNe are smaller than optical sizes. Their optical images show bipolar and elliptical morphology. It is not clear whether radio emission fills all the volume of the hot bubble or exists on the boundary of the hot bubble with the dense shell.

\begin{figure}[h]
\centering
\includegraphics[width=\columnwidth]{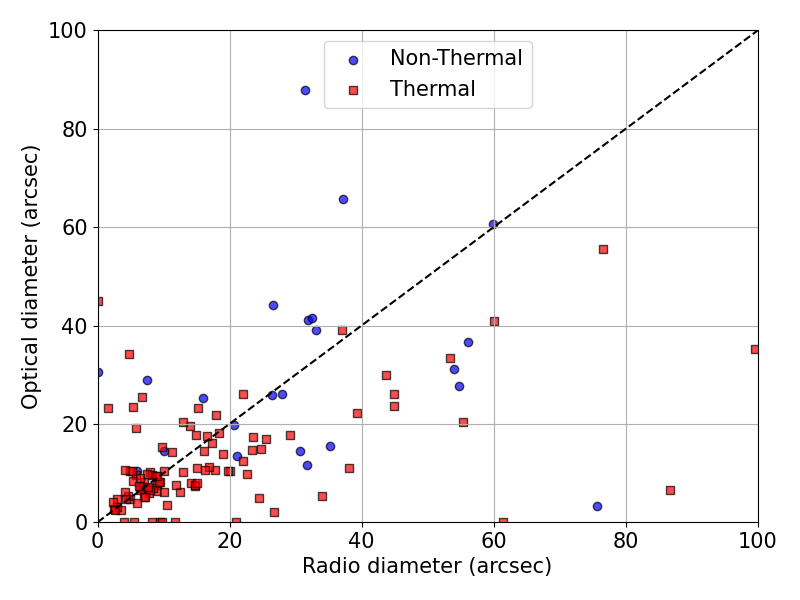}
\caption{Sizes of PNe dominated by thermal and non-thermal emission.}\label{diameters}
\end{figure}

\begin{figure}[h]
\centering
\includegraphics[width=\columnwidth]{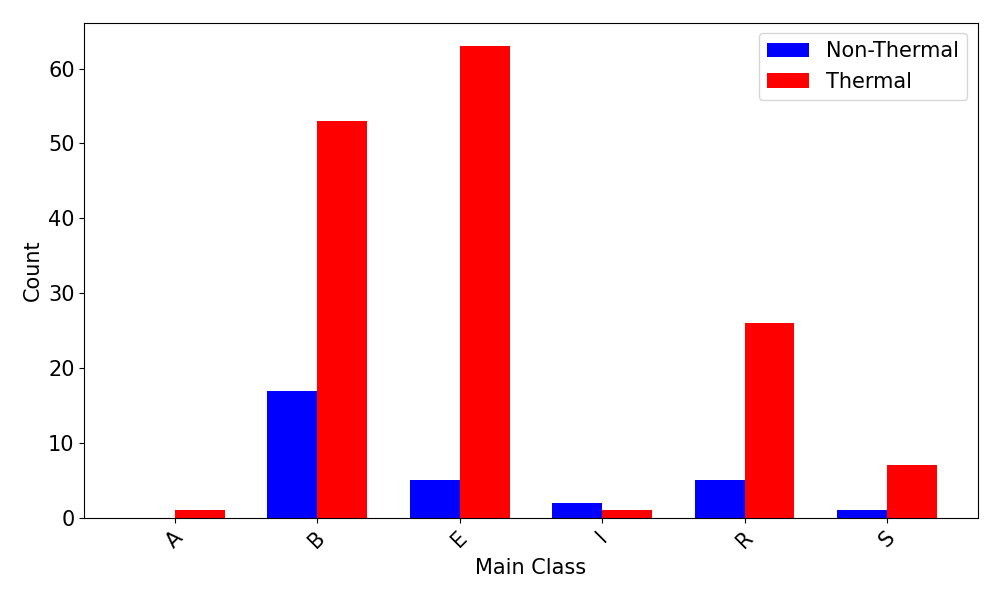}
\caption{Morphology of PNe dominated by thermal and non-thermal emission. R - round, B - bipolar, E - elliptical, I - irregular.}\label{shapes}
\end{figure}

\begin{figure}[h]
\centering
\includegraphics[width=\columnwidth]{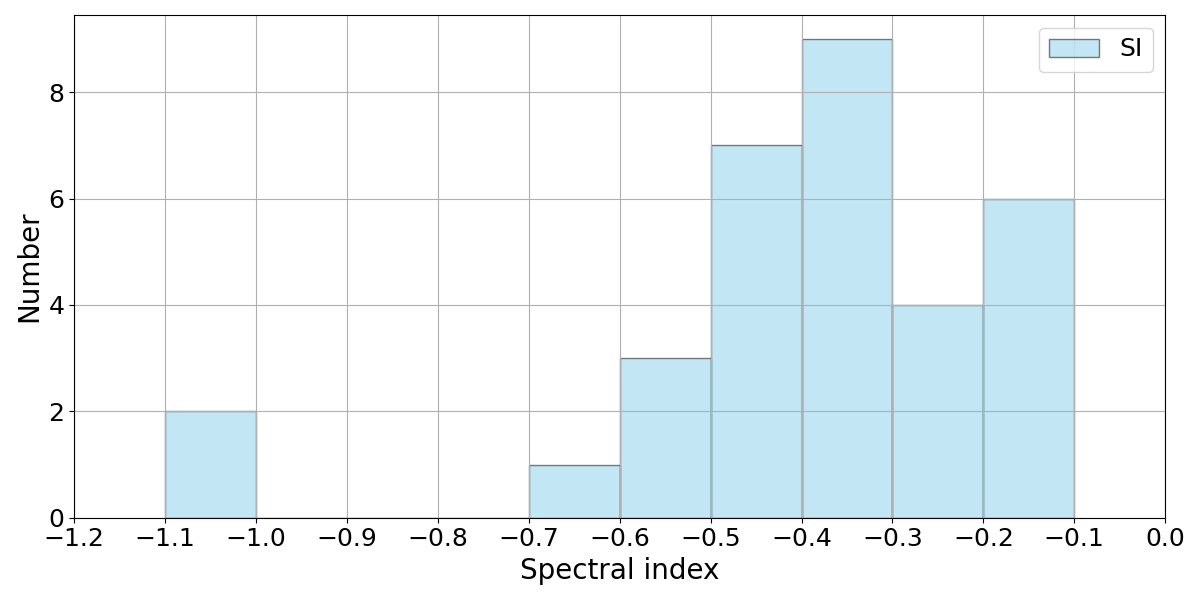}
\caption{Spectral indices of the non-thermal emission in PNe.}\label{si}
\end{figure}

A few sources clearly show bipolar radio emission: Abell\,53, AGP\,2, Kn\,7, M\,1-79, and PaEk\,1. Four of these PNe are classified in $\mathrm{H}\alpha$ as bipolar by \citet{2016JPhCS.728c2008P}. Abell\,53 is classified as a round PN. Their emission could trace the bipolar stellar outflow.

Finally, some of the PNe reveal patchy emission, e.g., Abell\,3, KK\,26, Kn\,43. These may be produced with the third mechanism discussed, the interaction of the wind with dense knots.

Non-thermal PNe have relatively large sizes. Most of them are larger than 20 arcsec (Figure\,\ref{diameters}). This may result from observational selection. One reason may be that smaller PNe are optically thick and non-thermal emission may be suppressed by an optically thick outer shell. Another reason is that in sources unresolved by LOFAR we cannot resolve non-thermal and thermal components spatially. Radio spectra may be dominated by the thermal component, but it does not exclude the contribution of non-thermal emission.

Most of PNe showing non-thermal emission are classified as bipolar (56\% compared to 35\% in thermally emitting sample) and only a few have elliptical shapes (17\% compared to 41\%). Figure\,\ref{shapes} shows the distribution of shapes for nonthermal and thermal sources with diameters $\geq 20 \, \rm{arcsec}$. However, in some cases, the morphological type assigned to a PN in optical may be wrong since the classification is based on the projected surface brightness distribution. For morphology determination, spatio-kinematical reconstruction is necessary in some cases. The deprojected structure of the observed radio emission is also not known. 

The spectral indices of most PNe cover the range $-0.1$ and $-0.7$, with two cases having values of $-1$ (Figure\,\ref{si}). The spectral index of non-thermal emission depends on the distribution of electron energies.
$\mathrm{N_e} = \mathrm{K E^{-\rho} dE}$
where $\mathrm{N_e}$ is a number of electrons per unit volume. Taking into account that 
$I_\nu \propto \nu^{\frac{-(\rho - 1)}{2}}$
radio spectra would indicate different electron distributions, e.g. $\rho = 2.4$ for the spectral index of $-0.7$ and $\rho = 1.8$ for the spectral index of $-0.4$. However, the observed spectral index may be affected by contribution from thermal emission in cases where radio observations cover a narrow range of frequencies. 

So far, magnetic fields have been observed in a small sample of PNe. \citet{2002A&A...392L...1G} discovered magnetic fields in NGC\,7027 and \citet{2009ApJ...695..930G} in a PN K\,3-35. \citet{2007MNRAS.376..378S} discovered toroidal fields in four PNe trace polarization of the dust. Non-thermal radio emission in IRAS\,15103-5754 \citep{2015ApJ...806..105S} and Sakurai's Object \citep{2024A&A...688L..21H} are linked with mass loss processes and wind interactions. In H\,{\sc ii} regions, which closely resemble PNe, non-thermal emission originates from the regions where electrons are accelerated in shocks \citep{2019A&A...630A..72P}.

\section{Summary}

New low-frequency observations largely increased the number of PNe with observed non-thermal emission and magnetic fields. We are not able to propose one common mechanism explaining non-thermal emission in all the observed PNe. However, non-thermal emission is more frequently observed in bipolar PNe, which indicates that mechanism that produces bipolar PNe is also responsible for formation of magnetic fields. The detected magnetic fields may be (circum-)stellar magnetic fields dragged out by the winds. Bipolar PNe produced in binary interaction more often possess structures such as low-ionisation knots or filaments which may interact with stellar wind \citep{2009A&A...505..249M}.

\section{Data availability}

The figures and fluxes of the 30 PNe are available at \href{https://zenodo.org/records/17377591?token=eyJhbGciOiJIUzUxMiJ9.eyJpZCI6ImQwMjQzYzBiLWUwMzctNGJhNy04NGFmLTBmZjkzNTY0NmM1NiIsImRhdGEiOnt9LCJyYW5kb20iOiIwOWU0N2NlODAxZmY1OWY2NGY1YjI1NjdlMjEwNzkzMCJ9.ey6XU5dTsWFh_lr1EbpZ_k1Hz6S52OqLE7YQoG7YRrF4u1LqkGD4qnPQZt2MCeN9cxV9y7Rp7EM5n_Cfx7Jicw}{this link}. The source catalog and images will be included in the official release of the LoTSS DR3 catalogue (Shimwell et al., in prep.).

\begin{acknowledgements}

RJD acknowledges support by the BMFTR ErUM-Pro program. We acknowledge Harish Vedantham for his comments on the manuscript. This research has made use of "Aladin sky atlas" developed at CDS, Strasbourg Observatory, France.

\end{acknowledgements}

\bibliographystyle{aa} 
\bibliography{aa} 

\end{document}